\documentclass[aps,prd,twocolumn,amsmath,amssymb,showpacs,nofootinbib]{revtex4}

\usepackage{graphicx,slashed,cancel,bbm,hyperref,framed,mathrsfs}
\usepackage[all]{xy}

\newcommand{\be}{\begin{equation}}
\newcommand{\ee}{\end{equation}}

\newcommand{\Det}{\mathrm{Det}}
\newcommand{\tr}{\mathrm{tr}}
\newcommand{\D}{\mathrm{d}}
\newcommand{\E}{\mathrm{e}}
\newcommand{\I}{\mathrm{i}}

\newcommand{\Tr}{\mathrm{Tr}}
\newcommand{\Lag}{\mathfrak{L}}

\newcommand{\V}{{\mathcal{V}}}

\newcommand{\Z}{{\mathcal{Z}}}

\renewcommand{\O}{{\mathcal{O}}}
\newcommand{\N}{{\mathcal{N}}}
\newcommand{\PO}{{\mathbbm{P}}}
\newcommand{\g}{\mathfrak{g}}
\newcommand{\fg}{\bar{\mathfrak{g}}}

\newcommand{\sfrac}[2]{{\textstyle\frac{#1}{#2}}}
\newcommand{\ul}{\underline}

\newcommand{\four}{\circ}%{4}
\newcommand{\five}{\bullet}%{5}

\newcommand{\opt}{\mathfrak}

\begin{document}
\title{Fermions in worldline holography}
\author{Dennis D.~Dietrich}
\author{Adrian Koenigstein}
\affiliation{Institut f\"ur Theoretische Physik, Goethe-Universit\"at, Frankfurt am Main, Germany}
\begin{abstract}
We analyse the worldline holographic framework for fermions.
Worldline holography is based on the observation that in the worldline approach to quantum field theory, sources of a quantum field theory over Mink$_4$ naturally form a field theory over AdS$_5$ to all orders in the elementary fields and in the sources. Schwinger's proper time of the worldline formalism automatically appears with the physical four spacetime dimensions in an AdS$_5$ geometry. The worldline holographic effective action in general and the proper-time profiles of the sources in particular solve a renormalisation group equation. By taking into account sources up to spin one, we reconstruct seminal holographic models. Considering spin two confirms AdS$_5$ as consistent background.
\end{abstract}
\pacs{
11.25.Tq %Gauge/string duality
12.40.Yx %Hadron mass models and calculations
11.10.Gh %Renormalization
11.10.Hi %Renormalization group flow
}
\maketitle
\section{Introduction}
Strong interactions offer an immensely rich phenomenology. Most of the time they overtax the computational abilities of the day and thus motivate us to put more energy into the development of new methods. In this context, since a few decades, the holographic idea \cite{'tHooft:1973jz,Maldacena:1997re,Gubser:1998bc,Witten:1998qj}---including the AdS/CFT correspondence---promises progress, and, for example, has been applied to quantum chromodynamics (QCD) \cite{Erlich:2005qh,Karch:2006pv,Polchinski:2000uf}, extensions of the Standard Model \cite{Hong:2006si,Dietrich:2008ni}, condensed-matter physics \cite{Sachdev:2011wg}, and the Schwinger effect \cite{Sato:2013dwa,Gorsky:2001up,Dietrich:2014ala}. All concrete instances of such correspondences discovered to date, however, hold for theories with a particle content different from QCD. For the time being, extrapolated ``bottom-up'' AdS/QCD descriptions are considered and capture the hadron spectrum thought-provokingly
accurately \cite{Karch:2006pv,Da Rold:2005zs}.
Yet, they lack a derivation from first principles and this is the motivation for delving into the fundamental reasons for which such an approach could be tenable. Considerable work has gone into this aspect already \cite{deTeramond:2008ht,deTeramond:2013it,Cata:2006ak}.

We managed to show \cite{Dietrich:2014ala,Dietrich:2016fby,Dietrich:2013kza,Dietrich:2013vla,Dietrich:2015oba,Dietrich:2015dka}, that a quantum field theory over Mink$_4$ readily turns into a field theory for its sources over AdS$_5$ in the framework of the worldline formalism \cite{Feynman:1950ir,Strassler:1992zr,Schubert:2001he} for quantum field theory. Schwinger's proper time naturally takes the role of the fifth dimension of the AdS$_5$ geometry\footnote{These worldline dualites are also available for different pairs of spaces including the non-relativistic case \cite{Dietrich:2013kza,Dietrich:2015oba,Dietrich:2015dka}.}. Schwinger's proper time sets a length scale (inverse energy scale), and this is the interpretation of the fifth dimension in holography \cite{Erlich:2005qh,Karch:2006pv,Polchinski:2000uf,Maldacena:1997re,Gubser:1998bc,Witten:1998qj} as well. Divergences occurring in a theory necessitate regularisation. In the worldline formalism they are naturally taken care of by proper-time regularisation, i.e., the introduction of a minimal positive proper-time. This proper-time regularisation corresponds to the UV-brane regularisation \cite{Erlich:2005qh,Karch:2006pv,Polchinski:2000uf,Maldacena:1997re,Gubser:1998bc,Witten:1998qj}.

\cite{Dietrich:2016fby,Dietrich:2015oba} demonstrated how such an AdS$_5$ formulation comes about to all orders in the sources and the elementary fields---matter and gauge.
Analysing the consequences of regulator independence of worldline holography in \cite{Dietrich:2016fby} identified it as a 
renormalisation group framework. In fact, we can define \cite{Dietrich:2013kza} worldline holography as a variational solution to a Wilson (gradient) flow \cite{Luscher:2009eq} and, from there, by the exact same computational steps obtain the identical result. Consequently, worldline holography was regulator independent all along, but we can and will define it through this requirement, henceforth.

In the past we concentrated on scalar elementary matter, as it provides the least impeded view at the underlying structure of worldline holography, due to the minimal number of internal degrees of freedom. There, among other things, worldline holography maps a free scalar theory onto a theory of arbitrarily high spins on AdS$_{d+1}$ \cite{FierzPauli}, as was previously conjectured \cite{Sundborg:2000wp}.

Here we turn to fermionic elementary matter, especially, as it makes up the matter part of the Standard Model. Considering fermions does neither overturn the worldline formalism \cite{Strassler:1992zr,Schubert:2001he} nor, as we shall see below, the worldline holographic framework in any way, but makes its phenomenology richer.

In Sect.~\ref{sec:wlholo} we present worldline holography for fermionic elementary matter. In Sect.~\ref{sec:free} we provide explicit computations in the free case. We derive the worldline holographic answer for (pseudo) scalar and (axial) vector sources, which calls for a comparison with \cite{Erlich:2005qh,Karch:2006pv,Polchinski:2000uf}. In Sect.~\ref{sec:selfcon} we confirm the self-consistency of the AdS background within our framework. In Sect.~\ref{sec:qed} we use our framework to study the renormalisation of quantum electrodynamics (QED). Section \ref{sec:sum} summarises our findings. 

\section{Worldline holography\label{sec:wlholo}}

As introduction we present the general framework for fermions in worldline holography. We start with one massless fermion flavour and a vector source $V$ combined with the gauge field $G$ in the `covariant derivative' $\mathbbm{D}=\partial-\I\mathbbm{V}$, where $\mathbbm{V}=G+V$. The generating functional for vector-current correlators is given by 
\begin{align}
Z=\langle\E^w\rangle=
\int[\D G]\E^{w-\frac{i}{4e^2}\int\D^4x\, G_{\mu\nu}^2},
\label{eq:genfun}
\end{align}
where 
\begin{align}
\label{eq:functint}
w
&=
\ln\int[\D\psi][\D\bar\psi]\E^{i\int\D^4x\bar\psi\I\slashed{\mathbbm{D}}\psi}
=\\&=
\ln\Det\,\I\slashed{\mathbbm{D}}
=
\Tr\ln\I\slashed{\mathbbm{D}}
=
\frac{1}{2}\Tr\ln \slashed{\mathbbm{D}}^2,
\label{eq:fermion}
\end{align}
and $\slashed{\mathbbm{D}}=\gamma^\mu\mathbbm{D}_\mu$.\footnote{Throughout the manuscript, we omit field independent normalisation terms.} The $\gamma^\mu$ stand for the anticommuting Dirac matrices $\{\gamma^\mu,\gamma^\nu\}=2\eta^{\mu\nu}$, where $\eta^{\mu\nu}$ represents the flat (inverse) metric. $\gamma^5=\I\gamma^0\gamma^1\gamma^2\gamma^3$ anticommutes with all $\gamma^\mu$, $\{\gamma^\mu,\gamma^5\}=0$. 
The first step in the derivation of the worldline representation for this determinant is replacing the logarithm by an exponential proper-time integral representation \cite{Strassler:1992zr,Schubert:2001he},
\begin{align}
\ln\slashed{\mathbbm{D}}^2=-\int_{\varepsilon>0}^\infty\frac{\D T}{T}\E^{-T\slashed{\mathbbm{D}}^2},
\label{eq:lnpt}
\end{align}
where we introduced the regulator $\varepsilon>0$. 
This, however, requires an operator with positive definite spectrum, which $\slashed{\mathbbm{D}}$ is not. For this reason, we continue with the last version of \eqref{eq:fermion},
\begin{align}
\slashed{\mathbbm{D}}^2 = \mathbbm{D}^2 - \frac{\I}{2} \sigma^{\mu\nu} [\mathbbm{D}_\mu , \mathbbm{D}_\nu]_-,
\label{eq:dss}
\end{align} 
where $\sigma^{\mu\nu}=\frac\I2[\gamma^\mu,\gamma^\nu]_-$.
The first addend  in \eqref{eq:dss} corresponds to the kinetic operator for scalar elementary matter \cite{Strassler:1992zr,Schubert:2001he,Dietrich:2016fby,Dietrich:2015oba}, the second is an additional potential term, i.e., one without open derivatives, also referred to as spin factor. 
In the worldline formalism\footnote{The worldline form \eqref{eq:wl} of the functional determinant \eqref{eq:fermion} is the particle dual of the determinant's wave(-function or field) representation as Feynman functional integral \eqref{eq:functint} in the sense of the particle-wave duality.} \cite{Feynman:1950ir,Strassler:1992zr,Schubert:2001he} after a Wick rotation, $w$ can be expressed as \cite{Dietrich:2014ala,Dietrich:2016fby,Dietrich:2013kza,Dietrich:2013vla} 
\begin{align}
w
=
&\int \D^4x_0\int_{\varepsilon>0}^\infty\frac{\D T}{2T^3}
\,\Lag\equiv\iint_\varepsilon^\infty\D^5x\,\sqrt{g}
\,\Lag ,
\label{eq:wl}
\\
\Lag
=
&-\frac{\mathcal{N}}{(4\pi)^2}
\int_\mathrm{P}[\D y]
\label{eq:lag}\\\nonumber&\hskip 1.5cm
\tr_\gamma\PO\,\E^{-\int_0^T\D\tau\{\frac{\dot y^2}{4}+\I \dot y \cdot\mathbbm V(x_0+y)+\frac{\I}{2}\sigma^{\mu\nu} [\mathbbm{D}_\mu , \mathbbm{D}_\nu]_-\}} ,
\end{align}
where the line element for the five-dimensional metric $g$ reads
\be
\D s^2=g_{MN}\D x^M\D x^N{=}+\frac{\D T^2}{4T^2}+\frac{\D x_0\cdot\D x_0}{T}
\label{eq:patch}
\ee
and $\sqrt{g}$ represents the square root of the absolute of its determinant.
`$\cdot$' stands for the contraction with $\eta^{\mu\nu}$. The Wick rotation turns the Minkowski $\eta_{\mu\nu}$ in the convention mostly plus to the Euclidean all plus. Simultaneously, \eqref{eq:patch} turns from an AdS$_{4,1}$ (frequently simply referred to as AdS$_5$; the pair of indices indicates the metric signature) to an AdS$_{5,0}$ (also referred to as H$_5$ or EAdS$_5$) line element. The isometry group of the five-dimensional AdS space is the conformal group of the corresponding four-dimensional flat space. $T$ represents Schwinger's proper time. One factor of $T^{-1}$ came from exponentiating the logarithm in \eqref{eq:lnpt}, while another factor of $T^{-2}$ arose when taking the functional trace.
The Lagrangian density $\mathcal{L}$ consists of a the path integral over all closed paths over the proper-time interval $[0;T]$, i.e., with $x(0)=x(T)$, where $x(\tau)=x_0+y(\tau)$. The normalisation cancels the free part, $\mathcal{N}\int_\mathrm{P}[\D y]\E^{-\frac14\int_0^T\D\tau\dot y^2}=1$. The $\D^4x_0$ integral translates otherwise equivalent paths to every position in space. The translations are the zero modes of the kinetic operator $\partial_\tau^2$, where $\dot y\equiv\partial_\tau y$. Separating them from the rest of the path integral also serves to make momentum conservation manifest. The choice of the representant loop for each equivalence class modulo translations is conventional; the centre-of-mass convention, for example, is defined through $\int_0^T\D\tau y=0$ and the starting-point convention through $y(0)=0=y(T)$\footnote{For details and more intermediate steps please see \cite{Dietrich:2013kza,Dietrich:2015oba,Strassler:1992zr,Schubert:2001he}.\label{foot3}
}. $\tr_\gamma$ indicates that the finite-dimensional trace over the $\gamma$ matrices remains to be taken\footnote{We have chosen to retain the $\gamma$-matrix representation of the anticommutativity of the fermions. Alternatively, an antiperiodic integration over Grassmann variables can be used to this effect \cite{Strassler:1992zr,Schubert:2001he}.}.  $\PO$ signifies that the exponential is path ordered, which is required due to the non-commutative nature of the $\gamma$ matrices. With the path ordering already in place we can consider non-Abelian flavour ($f$) and colour ($c$) groups right away as well, $\tr_\gamma\rightarrow\tr_{\gamma,f,c}$.

A rewrite of the free kinetic term of the worldline action, $\int_0^T\D\tau\frac{(\partial_\tau y)^2}{4}=\int_0^1\D\hat\tau\frac{(\partial_{\hat\tau} y)^2}{4T}$, where $\hat\tau=\tau/T$, shows that small values of $T$ confine $y$ to short relative distances, i.e., to the UV regime. Therefore, the proper-time regularisation $T\ge\varepsilon>0$ is a UV regularisation and corresponds to the UV-brane regularisation in holography \cite{Karch:2006pv,Erlich:2005qh,Polchinski:2000uf}. 

\subsection{Volume elements}

Taking stock, in the worldline formalism, $w$ automatically takes the form of an action \eqref{eq:wl} over AdS$_5$. In $e^w$, however, there are all powers of $w$. Thus, we have to show that this also holds for all other contributions to $Z$. The $n$-th power is given by
\begin{align}
w^n
=&
\prod_{j=1}^n
\int\D^4x_j\int_{\varepsilon}^\infty\frac{\D T_j}{2T_j^3}
\,\Lag(x_j,T_j).
\end{align}
The source-free part only depends on the positions of the single contributions $\Lag(x_j,T_j)$ relative to one another. As before, we separate off an absolute coordinate $x_0=\opt{x_0}(\{x_j\})$. It can be chosen as any linear combination of the $x_j$, like the centre of mass $\frac{1}{n}\sum_{j=1}^n x_j$, for example. This splits the $4n$ integrations into $4$ over the absolute, $\D^4x_0$, and $4(n-1)$ over the relative coordinates, $\D^{4(n-1)}\Delta$,
\begin{align}\nonumber
&\int \prod_{j=1}^n\D^4x_j\int\D^4x_0\delta^{(4)}[x_0-\opt{x_0}(\{x_k\})]
=\\=
&\int\D^4x_0\int\D^{4(n-1)}\Delta.
\end{align}

Accordingly, we define an overall proper time $T=\opt{T}(\{T_j\})$ and proper-time fractions $t_j=T_j/T$. Without the introduction of additional and thus artificial dimensionfull scales, on dimensional grounds, we have always that $\opt{T}(\{T_j\})=T\times\opt{T}(\{t_j\})$. A choice symmetric under the pairwise exchange of the $T_j$ makes the corresponding symmetry of $w^n$ manifest from the beginning. (Otherwise one could and would have to use the corresponding symmetry of $w^n$ to make the symmetry visible again.) The sum of all individual proper times $\opt{T}=\sum_{j=1}^nT_j$ is the arguably simplest choice satisfying these requirements. Implementing this change of variables with the help of  
\begin{align}
1=\int\D T\,\delta[T-\opt{T}(\{T_j\})]\prod_{j=1}^n\Big[\int\D t_j\,\delta\Big(t_j-\frac{T_j}{T}\Big)\Big].
\label{eq:choco}
\end{align}
yields
\begin{align}\nonumber
&\prod_{j=1}^n\int_{\varepsilon}^\infty\frac{\D T_j}{2T_j^3}
=\\=
&\int\D T\Big[\prod_{j=1}^n\int_{\varepsilon}^\infty\frac{\D T_j}{2T_j^3}\int\D t_j\,\delta\Big(t_j-\frac{T_j}{T}\Big)\Big]\times\\\nonumber&\hskip 5cm\times
\delta[T-\opt{T}(\{T_l\})]
=\\=
&\int\frac{\D T}{2T^3}T^{-2(n-1)}\int_{\frac{\varepsilon}{T}}^\infty\Big(\prod_{j=1}^n\frac{\D t_j}{2t_j^3}\Big)2\delta[1-\opt{T}(\{t_l\})].
\end{align}
Reexpressing $w^n$ in terms of the new variables, we obtain
\begin{align}
w^n
=&
\int\D^4x_0\int\frac{\D T}{2T^3}\int\frac{\D^{4(n-1)}\Delta}{T^{2(n-1)}}
\int_{\frac{\varepsilon}{T}}^\infty\Big[\prod_{j=1}^n\frac{\D t_j}{2t_j^3}
\times\\\nonumber&\hskip1.5cm\times
\Lag(x_0+x_j-x_0,Tt_j)\Big]
2\delta[1-\opt{T}(\{t_l\})].
\end{align}
Here $x_j-x_0$ is a function only of the relative coordinates $\Delta$ and not of the absolute coordinate $x_0$. Finally, we convert to dimensionless relative coordinates $\hat\Delta=\Delta/\sqrt{T}$, such that 
\begin{align}
w^n
=&
\int\D^4x_0\int\frac{\D T}{2T^3}\int\D^{4(n-1)}\hat\Delta
\int_{\frac{\varepsilon}{T}}^\infty\Big[\prod_{j=1}^n\frac{\D t_j}{2t_j^3}
\times\\\nonumber&\hskip0.9cm\times
\Lag(x_0+\widehat{x_j-x_0}\sqrt{T},Tt_j)\Big]
2\delta[1-\opt{T}(\{t_l\})],
\end{align}
which shows that $w^n$ takes the form of a Lagrangian density integrated over AdS$_5$ for all $n$.

\subsection{Contractions}

In order to be a genuine action over AdS$_5$ the contractions of all spacetime indices have to be performed with (inverse) AdS metrics. We start demonstrating this by extracting the dependence on ${x_j-x_0}$ and $y_j$ from the sources by means of a translation operator,
\begin{align}
\mathbbm{V}(y_j+x_j)
=%\\&=
\E^{(y_j+{x_j-x_0})\cdot\partial_{x_0}}\mathbbm{V}(x_0).
\end{align}
The relation holds for any function of $y_j+x_j$, i.e., here the vector $\mathbbm{V}_\mu$ but also the field tensor $\I[\mathbbm{D}_\mu,\mathbbm{D}_\nu]_-$. The combinations $x_j-x_0$ depend only on the relative coordinates $\Delta$. Putting everything into the Lagrangian density yields
\begin{align}\nonumber
&\Lag(x_j,T_j)
=
\Lag[x_0+(x_j-x_0),Tt_j]
=\\={}&
-\frac{{\mathcal{N}}}{(4\pi)^2}
\int_\mathrm{P}[\D y_j]\E^{-\int_0^1\D\hat\tau_j\frac{(\partial_{\hat\tau_j}y_j^\mu)g_{\mu\nu}(\partial_{\hat\tau_j}y_j^\nu)}{4t_j}}
\nonumber\times\\&\times
\tr_{\gamma,f,c}\PO\,\exp\Big(-\int_0^1\D\hat\tau_j\E^{[y_j^\lambda+(x_j-x_0)^\lambda]\cdot\frac{\partial}{\partial x_0^\lambda}}
\times\\&\times
\{\I(\partial_{\hat\tau_j}y_j^\rho)\mathbbm V_\rho(x_0)+\sfrac{\I}{2}t_j\check\sigma^{\rho\sigma}[\mathbbm{D}_{\rho}(x_0),\mathbbm{D}_{\sigma}(x_0)]_-\}\Big),
\nonumber
\end{align}
where $\hat\tau_j=\tau_j/T_j=\tau_j/(Tt_j)$.
The width of the $[\D y]$ integration is set by the four-dimensional part of the metric $g$ \eqref{eq:patch}. Consistently, the normalisation $\mathcal{N}$ compensates for the volume elements $\sqrt{g^{(4)}}$, the absolute of the determinant of the four-dimensional part of $g$. Consequently, after carrying out the $[\D y]$ integration, every pair of $y^\mu$-s generates an inverse metric $y^\mu_j y^\nu_k\xrightarrow{\int[\D y]} g^{\mu\nu}$ times a function of the proper times $\tau_j$ and $\tau_k$, which are integrated out subsequently. 
The $\D\Delta$ integration has a flat measure and the consistent non-unit volume element,
\begin{align}
\int\D^{4(n-1)}\hat\Delta=\int\frac{\D^{4(n-1)}\Delta}{T^{2(n-1)}}=\int\D^{4(n-1)}\Delta\big(\sqrt{g^{(4)}}\big)^{n-1}.
\end{align}
$g^{(4)}_{\mu\nu}$ alone describes a flat space, but has a non-unit normalisation. Accordingly, we introduced $\gamma$ matrices with the same normalisation, $\check\gamma^\mu=\sqrt{T}\gamma^\mu$, and therefore, $\{\check\gamma^\mu,\check\gamma^\nu\}=2g^{\mu\nu}$ as well as $\check\sigma^{\mu\nu}=T\sigma^{\mu\nu}$. All the above taken together shows, that the last two lines in
\begin{align}
w^n
=&
\int\frac{\D T}{2T}\int_{\frac{\varepsilon}{T}}^\infty\Big(\prod_{j=1}^n\frac{\D t_j}{2t_j^3}\Big)2\delta[1-\opt{T}(\{t_l\})]
\times\nonumber\\\nonumber&\times
\int\D^4x_0\sqrt{g^{(4)}}\int\D^{4(n-1)}\Delta\big(\sqrt{g^{(4)}}\big)^{n-1}
\times\\&\times
\prod_{k=1}^n\Lag[x_0+(x_k-x_0),Tt_k],
\end{align}
belong to a field theory over the space with the metric $g_{\mu\nu}^{(4)}$. ($T$ and the $t_j$ can be seen as external parameters in the four-dimensional context, and $T$ is exclusively present in the metric $g$.) Consequently, after carrying out the $[\D y]$ and $\D\Delta$ integrations as well as the trace over the $\gamma$ matrices, the remaining coordinate will be $x_0$ and all spacetime indices will be contracted with (inverse) metrics $g$ \cite{Dietrich:2016fby,Dietrich:2015oba}.

Integrating out the gauge field $G$ does not change this, as we can identically rewrite the action in the exponent of the integration measure using $g$ instead of $\eta$,
\begin{align}
\int\D^4x\sqrt{\eta}\eta^{\mu\kappa}\eta^{\nu\lambda} G_{\mu\nu}G_{\kappa\lambda}
=
\int\D^4x\sqrt{g^{(4)}}g^{\mu\kappa}g^{\nu\lambda} G_{\mu\nu}G_{\kappa\lambda},
\end{align}
which holds already at the level of the integrand.
Consequently, after also the $[\D G]$ integration is carried out, all contractions are still with $g$, which accounts for all powers of $T$. In sum, \eqref{eq:genfun} can be expressed as an action over AdS$_5$ for its sources to all orders and to all orders in the elementary fields.

\subsection{Fifth-dimensional components and renormalisation\label{sec:fifth}}

As explained in \cite{Dietrich:2016fby}, asking for the independence from the unphysical value of the UV regulator corresponds to a Wilson-Polchinsky renormalisation condition \cite{Wilson:1971bg} and is achieved by completing the five-dimensional field theory. Here we compile the essentials.
An expansion of the effective action in powers of gradients and sources after carrying out the $[\D y]$, $\hat\Delta$, $t_j$, and $\hat\tau_j$ integrations yields symbolically\footnote{\label{foot1}
The interaction part in \eqref{eq:lag} consists of a Wilson line part and a field tensor part. Therefore, it is locally invariant under the flavour transformation $V^\mu\rightarrow\Omega[V^\mu+\I\Omega^\dagger(\partial^\mu\Omega)]\Omega^\dagger$, which brings about hidden local symmetry \cite{Bando:1984ej}. Consequently, $Z_\varepsilon$ can also be expressed purely in terms of covariant derivatives \cite{Shifman:1980ui,Schmidt:1993rk},
$$
Z_\varepsilon=\iint_\varepsilon^\infty\D^5x\sqrt{g}\sum_n \#_n (g^{\four\four})^n(D_\four)^{2n},
$$
where $D=\partial-\I V$.
Furthermore, the proper-time regularisation preserves this symmetry; not so a momentum cutoff, for example. 
},
\begin{align}
\label{eq:genfuncser}
Z_\varepsilon
=
\int\D^4x_0\int_\varepsilon^\infty&\D T\sqrt{g}\sum_{n_\partial,n_V} \#_{n_\partial,n_V}
\times\\&~~~\times\nonumber
 (g^{\four\four})^\frac{n_\partial+n_V}{2}(\partial_\four)^{n_\partial}[V_\four(x_0)]^{n_V}.
\end{align}
The addends are to represent all possible occurring combinations. Each of the derivatives, generally, only acts on some of the sources and never to the right of the sources. The $\#_{n_\partial,n_V}$ are dimensionless numerical coefficients\footnote{There are only contributions from $n_\partial+n_V$ even.}. `$\four$' signifies that only four-dimensional contractions are carried out. 

$Z_\varepsilon$ depends on the proper-time regulator $\varepsilon>0$, the value of which, however, has a priori no physical meaning. Consequently, the physical effective action $\ln Z^\mathrm{phys}_\varepsilon$ should be regulator independent, i.e., we are looking for a solution to
\begin{equation}
\varepsilon\partial_\varepsilon \ln Z^\mathrm{phys}_\varepsilon\overset{!}{=}0,
\label{eq:renorm}
\end{equation}
which is a Wilson-Polchinsky renormalisation condition.

\eqref{eq:genfuncser} is already an action over AdS$_5$ \cite{Dietrich:2014ala,Dietrich:2013kza,Dietrich:2013vla,Dietrich:2015oba,Dietrich:2016fby}, albeit without fifth-dimensional components. The group of isometries of AdS$_5$ are the conformal group over Mink$_4$, including the invariance under scale transformations. Scale invariance would make the value of $\varepsilon$ irrelevant. 
In order to have the AdS isometries at our disposal, we have to complete the field theory by the missing components. 

Then again, the original four-dimensional theory has no fifth-dimensional polarisations. We can only remove them again if the $\breve\V_T=0$ is an allowed gauge condition. That means that the five-dimensional extension---in which the value of $\varepsilon$ is irrelevant---has to have five-dimensional local invariance under the flavor group. This fixes the form of the five-dimensional completion. Since $Z_\varepsilon$ is already locally invariant under four-dimensional transformations\footnote{This is manifest in the expansion shown in footnote \ref{foot1}. There the completion would proceed by replacing all flavour covariant derivatives by flavour and generally covariant derivatives.}
\begin{align}
\label{eq:calz}
\Z
=
\iint_\varepsilon^\infty\D^5x\sqrt{g}&\sum_{n_\partial,n_V}\#_{n_\partial,n_V}
\times\\&\times
 (g^{\five\five})^\frac{n_\partial+n_V}{2}(\nabla_\five)^{n_\partial}[\V_\five(x_0,T)]^{n_V},
\nonumber
\end{align}
is invariant under five-dimensional transformations. Here `$\five$' stands for the five-dimensional contractions, and $\nabla$ for the AdS covariant derivative.
\eqref{eq:calz} features the full AdS$_5$ isometries including scale invariance. Consequently, it is independent from the value of $\varepsilon$, if $\V(x_0,T)$ transforms like a five-dimensional vector. 
($\V$ does {\sl not} depend explicitly on $\varepsilon$.)
Imposing $\V_T=0$ gauge at the level of the action, would still manifestly preserves scale invariance, because scale transformations do not mix the tensor components. (The full symmetry would also be intact, but modulo a local flavour transformation.)

So far, $\Z$ is, however, only some functional of just any source configuration $\V$. It is a variational principle that makes it the (effective) action of a field theory, thus singling out a special field configuration (or configurations) as saddle point(s), $\breve\V$.
Through the boundary condition
\begin{align}
\breve\V_\mu(x_0,T=\varepsilon)=V_\mu(x_0)
\label{eq:inicond}
\end{align}
the four-dimensional polarisations are handed on to the five-dimensional solution $\breve\V$, and the normalisation as the source for exactly {\sl once} the vector current is preserved.\footnote{A quantised version of $\ln\Z$ also bears the necessary isometries to be a solution of the renormalisation condition \eqref{eq:renorm}, barring anomalies. The saddle point then is the leading contribution. The subleading correction is the fluctuation determinant. (The worldline formalism also relates this to the Gutzwiller trace formula \cite{Dietrich:2007vw,Gutzwiller:1971fy}, which also describes quantum systems through classical attributes.) In certain cases a distinction between `quantum' and `classical' turns out to be irrelevant, because the sum over the quantum contributions from all bulk fields cancels \cite{Giombi:2013fka}. 
} 

Moreover, \eqref{eq:inicond} coincides with the previous findings of worldline holography \cite{Dietrich:2014ala,Dietrich:2013kza,Dietrich:2015oba}, i.e., that the worldline formalism induces a Wilson (gradient) flow of the sources in the fifth dimension with this boundary condition.

Furthermore, \eqref{eq:inicond} localises the bare source configuration at the UV end of the fifth dimension, i.e., at small values of the proper time $T$ corresponding to short four-dimensional distances. In conjunction with the requirement \eqref{eq:renorm} that the effective action do not depend on the unphysical value of the UV regulator $\varepsilon$
marks this as a Wilson-Polchinsky renormalisation condition \cite{Wilson:1971bg}.

Finally, holography is the concept of extrapolating the sources from its boundary value (from the UV brane) into the bulk, and the effective action for the four-dimensional side of the holographic duality is described by the five-dimensional action evaluated on its saddle point\cite{Karch:2006pv,Erlich:2005qh,Polchinski:2000uf,Maldacena:1997re,Gubser:1998bc,Witten:1998qj}. As a matter of fact, there the computational steps are well-nigh identical, albeit, in parts, with a different reasoning.

Taking all the above into account, the desired cutoff independent effective action is obtained by evaluating $\Z$ on the saddle point configuration with the boundary condition \eqref{eq:inicond} and in $\breve\V_T=0$ gauge,
\begin{align}
\breve\Z
=
\iint_\varepsilon^\infty\D^5x\sqrt{g}&\sum_{n_\partial,n_V}\#_{n_\partial,n_V}
\times\\\nonumber&\times
 (g^{\five\five})^\frac{n_\partial+n_V}{2}(\nabla_\five)^{n_\partial}[\breve\V_\four(x_0,T)]^{n_V}.
\end{align} 
Hence, worldline holography identifies Schwinger's proper time as the fifth dimension \cite{Dietrich:2014ala,Dietrich:2016fby,Dietrich:2013kza,Dietrich:2013vla,Dietrich:2015oba} and fixes the fifth-dimensional profile of the sources as solution to the renormalisation group equation \eqref{eq:renorm}.

\section{Free case\label{sec:free}}

In order to obtain more insight into the worldline holographic formalism, let us turn to the free case. To this end, we switch off the coupling to the gauge bosons $G$ in \eqref{eq:genfun} by setting to zero the coupling to the gauge bosons $G$, which is tantamount to analysing $w$ with $\mathbbm{V}\rightarrow V$.\footnote{In low-energy scattering processes these should actually also be the kinematically dominant diagrams, justified by the observation that there the contributions with the lowest number of exchanged gauge bosons dominate \cite{Okubo:1963fa,Shifman:1978bx,De Rujula:1975ge,Dietrich:2012un}.}

For the sake of clarity, above we studied the vector, a rank-one source. In an expansion in the rank of the sources, however, we would thereby have omitted several other sources, namely the scalar $S$, the pseudoscalar $P$, and the axial vector $A$. These are also the sources needed to for a comparison to other holographic frameworks \cite{Karch:2006pv,Erlich:2005qh,Polchinski:2000uf,Hong:2006si,Dietrich:2008ni}.

With those sources in place
\begin{align}
w=\Tr\ln(i\slashed\partial+\Gamma),
\label{eq:wgamma}
\end{align}
where
\begin{align}
\Gamma=\slashed V+\gamma^5\slashed A+S+i\gamma^5P.
\label{eq:Gamma}
\end{align}
We again would like to use \eqref{eq:lnpt} for which we need an operator with positive definite spectrum. We choose the approach
\begin{align}
\O=(\O\O)^{1/2}=[(\O/\O^\dagger)(\O^\dagger\O)]^{1/2}
\end{align}
such that
\begin{align}
\Tr\ln\O=\frac{1}{2}\Tr\ln(\O\O^\dagger)+\frac{1}{2}\Tr\ln({\O}/{\O^\dagger}).
\label{eq:split}
\end{align}
In what follows, we are going to analyse $w$ maximally up to the fourth order in the fields and/or gradients. For $\mathcal{O}=\I\slashed\partial+\Gamma$, this result does not contain terms with an odd number of $\gamma^5$ matrices, which would come from the second addend in the previous equation \cite{DHoker:1995aat,Mondragon:1995ab}. (See also Appendix \ref{gfive}.) Hence, we only retain the first term,
\begin{align}
Z_\varepsilon
\supset&
-\frac{1}{(4\pi)^2}\iint_\varepsilon^\infty\D^5x\sqrt{g}\N\int_\mathrm{P}[\D y]\tr_{f,\gamma}\PO
\\
&\exp\Big[-\int_0^T\D\tau\Big(\frac{\dot y^2}{4}+
\gamma_\mathrm{R}\{\I\dot y^\mu L_{\mu}+\sfrac{1}{2}\sigma^{\mu\nu}L_{\mu\nu}
+\nonumber\\&\hskip 5mm+
\Phi\Phi^\dagger-\gamma^\mu D_\mu\Phi\}+\gamma_\mathrm{L}\{L\leftrightarrow R~\&~\Phi\leftrightarrow\Phi^\dagger\}\Big)\Big].
\nonumber
\end{align}
Here we switched to the basis
\begin{align}
L&=V+A,\\
R&=V-A,\\
\gamma_{L/R}&=(1\mp\gamma^5)/2,\\
\Phi&=S+\I P,
\end{align}
and introduced a flavour-covariant derivative
\begin{align}
{D}_\mu\Phi=\partial_\mu\Phi-\I L_{\mu}\Phi+\I\Phi R_{\mu}.
\label{eq:covder}
\end{align}
$L_{\mu\nu}$ ($R_{\mu\nu}$) stands for the field tensor for $L_\mu$ ($R_\mu$). 
The expansion to the level of the (flavour-covariant) kinetic terms for all source fields yields 
\begin{align}\label{eq:elsm}
&Z_\varepsilon
\supset-\frac{4}{(4\pi)^2}
\times\\\times
{}&\iint_\varepsilon^\infty\D^5x\sqrt{g}\,\tr_f[
\sfrac{1}{2}g^{\mu\nu}({D}_\mu\sqrt{T}\Phi)^\dagger({D}_\nu\sqrt{T}\Phi)
-|\sqrt{T}\Phi|^2
\nonumber\\
&\hskip 3cm+\sfrac{1}{12}g^{\mu\kappa}g^{\nu\lambda}(L_{\mu\nu}L_{\kappa\lambda}
+R_{\mu\nu}R_{\kappa\lambda})].
\nonumber
\end{align}
after carrying out the $[\D y]$ as well as $\D\tau$ integrations and dropping total derivatives.

According to the recipe detailed in Section \ref{sec:wlholo},
\begin{align}
\label{eq:Zholo}
&\Z
\supset-\frac{4}{(4\pi)^2}
\times\\\times
{}&\iint_\varepsilon^\infty\D^5x\sqrt{g}\,\tr_f[
\sfrac{1}{2}g^{MN}(\mathcal{D}_M\varPhi)^\dagger(\mathcal{D}_N\varPhi)
-\# |\varPhi|^2+
\nonumber\\
&\hskip 2cm+\sfrac{1}{12}g^{MK}g^{NJ}
(\mathcal{L}_{MN}\mathcal{L}_{KJ}+\mathcal{R}_{MN}\mathcal{R}_{KJ})],
\nonumber
\end{align}
back in Minkowski space and where $\sqrt{T}\Phi\rightarrow\varPhi$ \cite{Dietrich:2015oba}. The retention of the coefficients during the five-dimensional completion was needed to ensure the local invariance that allows us to gauge away the unphysical fifth polarisation. This holds for all fields but the spin-zero mass term, which is neither influenced by the introduction of fifth polarisations nor fifth gradients. Consequently, its coefficient is not thus protected. (In this context, it is important to remember that also in a purely four-dimensional utilisation of the worldline formalism additional conditions must be identified that are not automatically transferred by the formalism to ensure the correct renormalisation of the mass \cite{Ritus:1975,Dunne:2004nc}.) The way that $S$ is coupled to the elementary fermions in \eqref{eq:Gamma}, a finite mass $m$ of these fermions corresponds to a constant value $S=m$. Accordingly, in order to have a consistent framework, $\breve\varPhi=m\sqrt{T}$ must be an admissible classical solution for $\varPhi$ in \eqref{eq:Zholo}. For 4d homogeneous solutions, the classical equation of motion reads
\begin{align}
(2\partial_TT^{-1}\partial_T+ \# T^{-3})\breve\varPhi=0
\end{align}
and possesses power-law solutions $T^\alpha$, where $\alpha=1\pm\frac{1}{2}$ if $\#=3/2$. This is exactly the prediction of the holographic dictionary \cite{Gubser:1998bc,Witten:1998qj} for the fifth-dimensional mass of the scalar, which always includes the second independent solution $\propto T^{3/2}$.\footnote{This second solution is associated with spontaneous chiral symmetry breaking \cite{Erlich:2005qh} and thus should not contribute in the free case. The $T^{1/2}$ solution corresponds to the tachyon (squared) profile \cite{Bigazzi:2005md} for a free theory of elementary matter with the explicit mass $m$.} 
Consistently, inserting the solution $\propto T^{1/2}$ into the (flavour-covariant) spin-zero kinetic term generates a massterm for the axial vector $\propto m^2\mathcal{A}^2$, but not the vector \cite{Karch:2006pv,Erlich:2005qh}. For the free theory, the part of the action for the vector $\V$ does not contain any scale.

In order to see, what we can expect for an interacting gauge theory in its confining phase, let us represent the part of \eqref{eq:genfun} in which all sources are connected to a single matter loop by a confining term in the worldline action. (See also the discussion in \cite{Dietrich:2015oba}.) We can consider the Gaussian model from \cite{Dietrich:2013kza},
\begin{align}
S_\mathrm{Gauss}=\frac{1}{4}\int_0^T\D\tau(\dot y^2+c^2y^2),
\label{eq:Gauss}
\end{align}
or an area law for the area of the corresponding loop
\begin{align}
S_\mathrm{area}=\frac{1}{4}\int_0^T\D\tau\,\dot y^2+\mathrm{const.}\times\mathrm{area}.
\end{align}
As already discussed above, if the kinetic term sets the length scale, the typical length will be $O(\sqrt{T})$. Then $y^2$ as well as the area are $O(T)$, and, to logarithmic accuracy, we expect a (warp) factor $\E^{-\mathrm{const.}\!^\prime\times T}$ in the effective action.
Let us check our expectations for the first case \eqref{eq:Gauss}. Carrying out the path integral yields\footnote{There are other subleading differences between the effective warp factors of the different addends due to different powers of the modified worldline propagator on the ground floor.} 
\begin{align}
\mathcal{N}\int[\D y]\,\E^{-S_\mathrm{Gauss}}
&=
\prod_{n=1}^\infty\Big[1+\frac{c^2T^2}{4(2\pi)^2n^2}\Big]^{-d}
=\\&=
\Big[\frac{\sinh(cT/4)}{cT/4}\Big]^{-4}
=
\E^{-cT+O(\ln T)}.
\end{align}

Taking stock \eqref{eq:Zholo} with $\#=3/2$ and a confining potential/warp factor closely resembles the soft-wall model \cite{Karch:2006pv}. 

\subsection{Self-consistency of the AdS geometry\label{sec:selfcon}}

An effective action like \eqref{eq:genfuncser}, and particularly in the covariant form given in footnote \ref{foot1} can also be obtained for a generally curved background metric $\g$,
\begin{align}\nonumber
Z_\varepsilon
&=
\int_\varepsilon^\infty\frac{\D T}{2T^3}\int \D^4x_0\sqrt{\g}\sum_n \#_n (T\g^{\four\four})^n(\nabla_\four[\g])^{2n}
=\\&=
\iint_\varepsilon^\infty\D^5x\sqrt{\fg}\sum_n \sharp_n (\fg^{\four\four})^n(\nabla_\four[\g])^{2n},
\end{align}
where $\fg$ stands for the five-dimensional Fefferman-Graham \cite{Feffermann:1985} embedding of $\g$,
\begin{align}
\D s^2={\fg}_{MN}\D x^M\D x^N{=}\natural\Big(\frac{\D T^2}{4T^2}+\frac{\g_{\mu\nu}\D x^\mu\D x^\nu}{T}\Big),
\end{align}
$\sharp_n=\#_n\natural^{n-5/2}$, and $\nabla[\g]$ for the Levi-Civita connection. The $\#_n$ are the DeWitt-Gilkey-Seeley coefficients \cite{DeWitt:1965}.

As seen above, the independence \eqref{eq:renorm} from $\varepsilon$ can be achieved by means of the completion to a five-dimensional action 
\begin{align}
\Z
=
\iint_\varepsilon^\infty\D^5x\sqrt{\fg}\sum_n \sharp_n (\fg^{\five\five})^n(\nabla_\five[\fg])^{2n},
\end{align}
and its subsequent evaluation on its saddle point for the boundary condition 
\begin{align}
\breve\fg_{\mu\nu}(x_0,T=\varepsilon)=\frac{\natural}{\varepsilon}\g_{\mu\nu}(x_0)
\label{eq:gbound}
\end{align}
in the gauge where
\begin{align}
\breve\fg_{TT}\overset{!}{=}\natural g_{TT}~~~\&~~~\breve\fg_{T\nu}\overset{!}{=}0,
\label{eq:ggauge}
\end{align}
with $g_{TT}$ from \eqref{eq:patch}. This corresponds to the absence of deviations from $g$ with fifth-dimensional polarisations, 
\begin{align}
h_{TN}\overset{!}{=}0\,\forall N.
\end{align}

The two leading terms \cite{Shapiro:2008sf} correspond to a negative cosmological constant and an Einstein-Hilbert term,
\begin{align}
\Z\supset-\frac{1}{3(4\pi)^2}\iint_\varepsilon^\infty\D^5x\sqrt{\fg}(R[\fg]+12).
\label{eq:eh}
\end{align} 
As a consequence, the corresponding Einstein equations admit an AdS$_5$ solution with the squared AdS curvature radius 
\begin{align}
\natural=
\frac{(5-1)(5-2)}{12}=1.
\end{align}
Taking into account the boundary \eqref{eq:gbound} and gauge conditions \eqref{eq:ggauge} the solution is $\breve\g=g$.
Therefore, to this order, an AdS background is a self-consistent prediction of the formalism.

At higher orders, AdS, being a space of constant curvature, is still a saddle-point solution, although generally with a different curvature radius.
The AdS$_5$ isometry group does not depend on the value of the curvature radius and is always the conformal group over Mink$_4$. (Analogously, Mink$_4$ is Poincar\'e invariant for every value of the speed of light.) As a consequence, the value of the AdS radius is of secondary importance. For one thing,
\begin{align}
\Z
=
\iint_\varepsilon^\infty\D^5x\,\breve\fg^{1/2}&\sum_{n_\partial,n_V}\sharp_{n_\partial,n_V}
\times\\&\times
 (\breve\fg^{\five\five})^\frac{n_\partial+n_V}{2}(\nabla_\five[\breve\fg])^{n_\partial}[\V_\five(x_0,T)]^{n_V},
\nonumber
\end{align}
where $\sharp_{n_\partial,n_V}=\#_{n_\partial,n_V}\natural^{(n_\partial+n_V-5)/2}$, is identical to \eqref{eq:calz}, which itself does not depend on $\natural$. Likewise, the covariant derivatives are independent from the curvature radius as is the (1,3) Riemann tensor. 

\section{Holographic 2-loop charge renormalisation of QED\label{sec:qed}}

Interpreting the vector source $V$ as a (background) gauge field $Z_\varepsilon$ is the QED effective action in the background-field formalism. There is a logarithmic divergence in the leading term
\begin{align}
Z_\varepsilon=\#_{2,2}\iint_\varepsilon^\infty\D^5x\sqrt{g}g^{\mu\kappa}g^{\nu\lambda}V_{\mu\nu}V_{\kappa\lambda},
\label{eq:qedz}
\end{align}
where $V_{\mu\nu}$ represents the (presently Abelian) field-strength tensor. The divergence appears in the $\D T$ integration, where there is a factor of $T^{-3}$ from the volume element and two factors of $T$, one from each metric $g^{\mu\nu}$, which makes an overall $\D T/T$. To two loops 
for $N_f\times N_c$ quarks \cite{Schubert:2001he}, 
\begin{align}
\#_{2,2}=2\frac{N_fN_c}{(4\pi)^2}\Big(-\frac{1}{3}-\frac{e^2}{(4\pi)^2}\Big).
\end{align}
The full five-dimensional action
\begin{align}
\Z=\#_{2,2}\iint_\varepsilon^\infty\D^5x\sqrt{g}g^{MK}g^{NL}\V_{MN}\V_{KL},
\label{eq:fivetwo}
\end{align}
is independent from $\varepsilon$. Here capital indices run over all five dimensions. 
The corresponding saddle-point equations are given by
\begin{align}
g^{NL}\nabla_N\breve\V_{KL}=0.
\end{align}
In the axial gauge $\breve\V_T=0$ these equations of motion also imply Lorenz gauge $\partial\cdot\breve\V=0$.
Then the remaining transverse components (here in 4d momentum space) must obey
\begin{align}
\Big(\partial_T^2-\frac{p^2}{4T}\Big)\tilde{\breve\V}^\perp=0.
\end{align}
The normalisable solution with \eqref{eq:inicond} is given by
\begin{align}
\tilde{\breve\V}^\perp=\tilde V^\perp(p)\frac{\sqrt{p^2T}K_1(\sqrt{p^2T})}{\sqrt{p^2\varepsilon}K_1(\sqrt{p^2\varepsilon})},
\end{align}
where Bessel's $K_n$ is defined in 9.6.1.~ff.~in \cite{as}, and for which (see 9.6.28 in \cite{as})
\begin{align}
\partial_T\tilde{\breve\V}^\perp=\tilde V^\perp(p)\frac{p^2K_0(\sqrt{p^2T})/2}{\sqrt{p^2\varepsilon}K_1(\sqrt{p^2\varepsilon})}.
\end{align}
Putting this solution back into the 4d Fourier transformed action \eqref{eq:fivetwo} we obtain a surface term,
\begin{align}
\breve\Z
&=
4\#_{2,2}\int\D^4x_0\,\eta^{\nu\lambda}[\breve\V_\nu^\perp\partial_T\breve\V_\lambda^\perp]_\varepsilon^\infty
=\\&=
4\#_{2,2}\int\frac{\D^4p}{(2\pi)^4}\,\eta^{\nu\lambda}[\tilde{\breve\V}_\nu^\perp\partial_T\tilde{\breve\V}_\lambda^{\perp*}]_\varepsilon^\infty
=\\&=
-2\#_{2,2}\int\frac{\D^4p}{(2\pi)^4}\eta^{\nu\lambda}\tilde V_\nu^\perp\tilde V_\lambda^{\perp*}p^2K_0(\sqrt{p^2\varepsilon}),
\end{align}
where $\tilde{}$ marks the Fourier transform and $^*$ the complex conjugate.
Making use of 9.6.13.~from \cite{as},
\begin{align}
\breve\Z
&=
\#_{2,2}\int\frac{\D^4p}{(2\pi)^4}\underbrace{\eta^{\nu\lambda}\tilde V_\nu^\perp\tilde V_\lambda^{\perp*}p^2}_{\hat=|\tilde V_{\mu\nu}|^2/2}\{\ln(p^2\varepsilon)+O[(p^2\varepsilon)^0]\}.
\label{eq:div}
\end{align}
In our conventions, where the coupling $e$ is absorbed in the field, the prefactor of the kinetic term equals $-(4e^2)^{-1}$. 

To two loops, the $\beta$ function describing the running of the coupling with the scale $\mu$ is given by
\begin{align}
\frac{\D e}{\D\ln\mu}=\beta_1e^3+\beta_2e^5.
\label{eq:beta}
\end{align}
Integrating \eqref{eq:beta} and solving for $e^{-2}(\mu)-e^{-2}(\mu_0)$\footnote{As usual, we subtract the bare contribution from the induced term.} yields 
\begin{align}
e^{-2}(\mu)-e^{-2}(\mu_0)=-2(\beta_1+\beta_2e^2)\ln\frac{\mu}{\mu_0}+\dots,
\end{align}
where the ellipsis stands for terms of $O(e^4)$, and this order also depends on the three-loop coefficient. The comparison of the divergent pieces yields
\begin{align}
2\#_{2,2}\ln(p^2\varepsilon)=-e^{-2}=(\beta_1+\beta_2e^2)2\ln\frac{\mu}{\mu_0}.
\end{align}
Upon identification of $\ln(p^2\varepsilon)\leftrightarrow2\ln\frac{\mu}{\mu_0}$ we obtain
\begin{align}
\beta_1+\beta_2e^2=2\#_{2,2}=4\frac{N_fN_c}{(4\pi)^2}\Big(-\frac{1}{3}-\frac{e^2}{(4\pi)^2}\Big),
\end{align}
which are the known $\beta$-function coefficients.\footnote{In non-holographic worldline computations a finite mass was needed as IR regulator and its behaviour under renormalisation had to be determined in an additional computation to achieve this result \cite{Ritus:1975,Dunne:2004nc}. Here, we considered the massless case, chose the integrable, i.e., IR finite solution, and did not need any additional input.}

In the worldline formalism there are no subdivergences in the two-loop contribution \cite{Schubert:2001he} to the coefficient $\#_{2,2}$. (This does not only hold for proper-time regularisation, but also other four-dimensional regularisation schemes like Pauli-Villars.) The absence of subdivergences is known to persist for the quenched contributions to all loops \cite{Dunne:2004nc}. For higher unquenched orders the analysis is still pending.

Non-holographic renormalisation of QED was treated in the worldline formalism before \cite{Ritus:1975,Dunne:2004nc}. There, obtaining the two-loop term in the analog of \eqref{eq:div} required knowledge of the counter term from mass renormalisation, where the mass was used as infrared regulator. We here work with massless elementary matter. Hence, there is no mass renormalisation. Asking for the integrability of the saddle-point solution led to an infrared finite result and the known two-loop contribution.

~\\

We never forced $\varepsilon$ to be small. [\eqref{eq:div} only presents the behaviour of $\breve\Z$ for if $\varepsilon$ were small.] $\varepsilon$ was introduced to regularise the UV divergence of $Z$. Then, we had in mind to send the regulator to zero at the end of the calculation. At non-zero $\varepsilon$ the renormalisation condition \eqref{eq:renorm} makes $\varepsilon$ a scale. In case we would like to keep $\varepsilon$ in its original role as regulator, we can introduce counter terms for the divergent pieces. In \eqref{eq:div}, for instance,
\begin{align}
\breve\Z
={}&
\#_{2,2}\int\frac{\D^4p}{(2\pi)^4}\eta^{\nu\lambda}\tilde V_\nu^\perp\tilde V_\lambda^{\perp*}p^2
\times\\&\hskip 2cm\times
\{\ln(\mu^2\varepsilon)+\ln(p^2/\mu^2)+O[(p^2\varepsilon)^0]\},
\nonumber
\end{align}
the first addend inside the braces, which diverges when $\varepsilon\rightarrow0$, must be compensated by the introduction of a counter term which can also contain additional finite parts. Now $\mu^2$ plays the role of the scale and $\varepsilon$ remains the regulator. 

\section{Summary\label{sec:sum}}

We studied the worldline holographic framework for fermionic elementary matter. Worldline holography maps a $d$ dimensional quantum field theory onto a $d+1$ dimensional field theory for the sources of the former, to all orders in the elementary fields and sources. The $d+1$ dimensional metric is the Fefferman-Graham embedding \cite{Feffermann:1985} of the $d$ dimensional one. For Mink$_d$ this results in AdS$_{d+1}$. Worldline holography is the solution to a Wilson-Polchinski renormalisation condition \eqref{eq:renorm}, which guarantees the independence of physical quantities from the ultraviolet regulator. (Infrared scales can be handled analogously \cite{Dietrich:2016fby}.) As consistency check we holographically derived the QED beta-function coefficient to two loops, in Sect.~\ref{sec:qed}. In Sect.~\ref{sec:free}, we explicitly determine the worldline holographic dual for a fermionic field theory on Mink$_d$ with sources up to spin one and find a theory akin to the seminal holgraphic model \cite{Karch:2006pv}. Turning to spin two in Sect.~\ref{sec:selfcon} allows us to confirm that AdS$_{d+1}$ is a selfconsistent solution of the worldline holographic framework.

\section*{Acknowledgments}

The authors would like to thank
Ibrahim Akal,
Stan Brodsky,
Guy de T\'eramond,
Luigi Del Debbio,
Florian Divotgey,
Gerald Dunne,
Gia Dvali,
Joshua Erlich,
J\"urgen Eser,
Francesco Giacosa,
Holger Gies,
C\'esar G\'omez,
Stefan Hofmann,
Paul Hoyer,
Johannes Kirsch,
Sebastian Konopka,
Matti J\"arvinen,
Yaron Oz,
Stefan Rechenberger,
Dirk Rischke,
Ivo Sachs,
Andreas Sch\"afer,
Stefan Theisen,
and
Roman Zwicky
for discussions. 

\appendix

\section{$\gamma^5$-odd terms\label{gfive}}

By inspection of the source \eqref{eq:Gamma}, up to the fourth order in the fields and gradients taken together, one expects terms with one $\gamma^5$, $O[(\partial-\I V)^3A]$, and three $\gamma^5$, $O[(\partial-\I V)A^3]$, to occur and to be contained in the ${\O}/{\O^\dagger}$ part of \eqref{eq:split} \cite{DHoker:1995aat,Mondragon:1995ab}. [To this order, there are no corresponding contributions from the pseudoscalar $P$, as its $\gamma^5$ would have to be ballanced by four $\gamma^\mu$ from vectors and/or (an even number of) axial vectors, which would amount to order five.] These terms encode the axial anomaly of the theory \cite{DHoker:1995aat,Mondragon:1995ab}. 

Confusingly, also the ${\O}{\O^\dagger}$ part seems to yield such a contribution,
\begin{align}\label{eq:elsm}
&Z_\varepsilon
\supset\frac{1}{(8\pi)^2}
\iint_\varepsilon^\infty\D^5x\sqrt{g}\,T^2\epsilon^{\mu\nu\kappa\lambda}\tr_f(L_{\mu\nu}L_{\kappa\lambda}-R_{\mu\nu}R_{\kappa\lambda}),
\nonumber
\end{align}
contrary to the conclusions in \cite{DHoker:1995aat,Mondragon:1995ab}. Concentrating on the Abelian part, in a momentum-space computation the term drops out once momentum conservation is imposed. In \cite{Mondragon:1995ab} this can be seen from their Feynman rules on the first line of Table 1. There, the trace over the products of the $\sigma_{\mu\nu}$ terms in the vertices yields an $\epsilon_{\mu\nu\kappa\lambda}$, which, however, is contracted with $p$ and $-p$ and thus vanishes for symmetry reasons. 

Independently, we have carried out a momentum-space Feynman-diagram computation, where we encounter
\begin{align}
\tr[\gamma_\mu\tilde\Gamma(p)\gamma_\nu\tilde\Gamma^*(p)]
\supset
-4i\epsilon_{\mu\kappa\nu\lambda}(\tilde A_\kappa\tilde V_\lambda^*-\tilde V_\kappa\tilde A_\lambda^*),
\end{align}
which is either contracted with $p^\mu p^\nu$ or $\eta^{\mu\nu}$ and thus vanishes in any case.

In coordinate space, though, also by inspection of (5) in \cite{Mondragon:1995ab}, one identifies again the combination
\begin{align}
\tr[(\mathcal{O}\mathcal{O}^\dagger)^2]
\supset
\tr(\gamma^5\sigma^{\mu\nu}\sigma^{\kappa\lambda}A_{\mu\nu}V_{\kappa\lambda})
\end{align}
up to a numerical prefactor. The authors of \cite{Mondragon:1995ab} show that the analogous contribution from $\mathcal{O}/\mathcal{O}^\dagger$ saturates the axial anomaly, such that $\mathcal{O}\mathcal{O}^\dagger$ should not contribute here.

~\\

In any case, these terms are purely topological, do not contribute to the equations of motion, and yield at most surface contributions. They are readily embedded in our five-dimensional setting. Using the five-dimensional Levi-Civita {\sl tensor}
\begin{align}
\mathcal{E}^{MNKLJ}=\epsilon^{MNKLJ}/\sqrt{g}=2T^3\epsilon^{MNKLJ}
\end{align}
and the $T$ components of the (diagonal) f\"unfbein,
\begin{align}
{E}_T^A=\delta_T^A\sqrt{g_{TT}}=\delta_T^A/2T,
\label{eq:5bein}
\end{align}
we get\footnote{The index $T$ stands for the fifth component. In particular, this means that we do {\sl not} sum over it, if it appears in pairs. We mark that by an underline when necessary.}
\begin{align}
\mathcal{E}^{MNKL\ul{T}}{E}_{\ul T}^A=T^2\epsilon^{\mu\nu\kappa\lambda}\delta^A_T
\label{eq:lct}
\end{align}
and can thus express all contractions in \eqref{eq:elsm} with five-dimensional objects\footnote{One could still contract \eqref{eq:lct} with a constant vector $n_A$.}.

Consistently, \eqref{eq:5bein} satisfies ${E}_M^A{E}_N^BH_{AB}=g_{MN}$, where $H_{AB}$ is a reference metric. [If chosen flat (and ``unity'') it gives rise to the last of the equalities in \eqref{eq:5bein}.] This is also in line with the expression for the CP-even term (only coming in at higher orders)
\begin{align}
&T^4(\epsilon^{\mu\nu\kappa\lambda}V_{\mu\nu}V_{\kappa\lambda})^2
\nonumber=\\=
{}&\mathcal{E}^{\mu\nu\kappa\lambda\ul T}g_{\ul T\ul T}\mathcal{E}^{\alpha\beta\gamma\delta\ul T}V_{\mu\nu}V_{\kappa\lambda}V_{\alpha\beta}V_{\gamma\delta}
\xrightarrow{5d}\\\xrightarrow{5d}
{}&\mathcal{E}^{MNKLJ}g_{JE}\mathcal{E}^{ABCDE}\V_{MN}\V_{KL}\V_{AB}\V_{CD}
=\\=
{}&4!\,g^{MA}g^{NB}g^{KC}g^{LD}\V_{MN}\V_{KL}\V_{[AB}\V_{CD]},
\end{align}
which ultimately can be expressed purely using (inverse) metrics. (In this passage, $V$ is a placeholder for $L$ or $R$, respectively.)

The introduction of the vielbein as ``square root'' of the metric is consistent, but its origin can be elucidated further.
Consider the five-dimensional Chern-Simons term \cite{Chern:1974ft}
\begin{align}
C\!S_5=\epsilon^{MNKLJ}\tr_f(&\V_{MN}\V_{KL}\V_J\nonumber
-\\&-\sfrac{1}{2}\V_{MN}\V_K\V_L\V_J\nonumber
+\\&+\sfrac{1}{10}\V_M\V_N\V_K\V_L\V_J).
\end{align}
Under local transformations $\V_M\rightarrow\Omega[\V_M+\I\Omega^\dagger(\partial_M\Omega)]\Omega^\dagger$ it changes only by a total derivative. We are, however, working on a manifold with boundary (at $T=\varepsilon$), which can render such otherwise cyclic components physical.\footnote{Another example is the zero mode of the temporal gauge field in thermal field theory \cite{Gies:1998vt}, where it is due to the compactification of the time direction.} The gauge transformation parametrises such degrees of freedom, the longitudinal components of $\V$. Making the latter explicitly visible using the St\"uckelberg trick one lifts them to auxiliary fields and looks whether and where they contribute. Here the St\"uckelberg trick amounts to
\begin{align}
\V_M\rightarrow\bar\V_M=\V_M+\partial_M\Sigma,
\end{align}
where the 5d longitudinal part $\partial_M\Sigma$ corresponds to $\I\Omega^\dagger(\partial_M\Omega)$.\footnote{Since $L_\mu/R_\mu$ couples to $\Phi$ only through the covariant derivative \eqref{eq:covder}, a linear combination of the longitudinal components $\Sigma_{L/R}$ and the spin-zero fields $\Phi$ is cyclic.} Then, perturbatively,
\begin{align}
C\!S_5[\bar\V]=C\!S_5[\V]+\epsilon^{MNKLJ}\tr_f[(\partial_M\Sigma)\V_{NK}\V_{LJ}]+\dots,
\end{align}
where the ellipsis stands for higher orders in $\partial_M\Sigma$. For configurations without fifth polarisations nor gradients, $C\!S_5$ vanishes exactly. To the contrary,
\begin{align}\nonumber
&\iint_\varepsilon^\infty\D^5x\,\sqrt{g}\,\mathcal{E}^{MNKLJ}\,\tr_f[(\partial_M\Sigma)\V_{NK}\V_{LJ}]
=\\=
-{}&\iint_\varepsilon^\infty\D^5x\,\epsilon^{MNKLJ}\partial_M\,\tr_f[\Sigma\V_{NK}\V_{LJ}]
=\\=
+{}&\int\D^4x\,\epsilon^{TNKLJ}\,\tr_f[\Sigma\V_{NK}\V_{LJ}]_{T=\varepsilon}
\xrightarrow{4d}\\\xrightarrow{4d}
+{}&\int\D^4x\,\underbrace{\epsilon^{T\nu\kappa\lambda\rho}}_{=\epsilon^{\nu\kappa\lambda\rho}}\,\tr_f[\Sigma(T=\varepsilon) V_{\nu\kappa}V_{\lambda\rho}]
\end{align}
is, in general, nonzero. The presence of the boundary makes the zero mode of $\Sigma$ physical. All other terms depend only on derivatives of $\Sigma$.  
This term encodes the chiral anomaly \cite{Bardeen:1969md}. Consistently, $C\!S_5$ also contains the Wess-Zumino-Witten term \cite{Wess:1971yu},
\begin{align}
&C\!S_5[\partial\Sigma]
=\\=
\nonumber{}&+\epsilon^{MNKLJ}\tr_f[(\partial_M\Sigma)(\partial_N\Sigma)(\partial_K\Sigma)(\partial_L\Sigma)(\partial_J\Sigma)]
=\\=
{}&-\epsilon^{MNKLJ}\partial_M\tr_f[\Sigma(\partial_N\Sigma)(\partial_K\Sigma)(\partial_L\Sigma)(\partial_J\Sigma)].
\end{align}

~\\

We end this appendix with an interesting observation. When doing the momentum-space Feynman-diagram computation, for instance, of the two-point function, when using the Feynman trick to exponentiate the denominators of the propagators one reproduces the fifth-dimensional structure of the worldline framework,
\begin{align}
&\int\frac{\D^4p}{(2\pi)^4}\tr\Big[\frac{\slashed p+\frac{\slashed q}{2}}{(p+\frac{q}{2})^{2}}\tilde\Gamma(q)\frac{\slashed p-\frac{\slashed q}{2}}{(p-\frac{q}{2})^{2}}\tilde\Gamma(-q)\Big]
=\\=&
\int\frac{\D^4p}{(2\pi)^4}\int_0^\infty\D T_1\D T_2\,\E^{{T_1(p+\frac{q}{2})^{2}+T_2(p-\frac{q}{2})^{2}}
}
\times\nonumber\\&\times
\tr\Big[\Big(\slashed p+\frac{\slashed q}{2}\Big)\tilde\Gamma(q)\Big(\slashed p-\frac{\slashed q}{2}\Big)\tilde\Gamma(-q)\Big]
=\\=&
\frac{1}{(4\pi)^2}\int_\varepsilon^\infty\frac{\D T}{T^2}\int_0^1\D\hat\tau\,T\,\E^{T\hat\tau(1-\hat\tau)q^2}T\hat\tau(1-\hat\tau)
\times\nonumber\\&\times
(-q^2{\eta^{\mu\nu}}/{2}-q^\mu q^\nu)\tr[\gamma_\mu\tilde\Gamma(q)\gamma_\nu\tilde\Gamma(-q)].
\end{align}
One recognises the sum of the Feyman parameters $T=T_1+T_2$ as Schwinger's proper time. The combination $T\hat\tau(1-\hat\tau)$ is known as worldline propagator \cite{Strassler:1992zr,Schubert:2001he}. $\int_0^1\D\hat\tau\E^{T\hat\tau(1-\hat\tau)q^2}$ is also the basic second-order form factor in the heat-kernel expansion in the background-field formalism \cite{Barvinsky:2015bky}. Hence, calling the present framework proper-time holography or even heat-kernel holography, would not seem amiss, but it was the worldline formalism that allowed us to make the connection to AdS$_5$ and renormalisation to all orders.

\end{document}